\documentclass[12pt]{iopart}
\usepackage{graphicx}
\begin{document}

\title[Partitioning of Min proteins between daughter cells after septation]
	{Modeling partitioning of Min proteins between daughter cells after septation in 
	{\em Escherichia coli}}
\author{Supratim Sengupta$^{1,2}$ and Andrew Rutenberg$^{1}$}

\address{$^1$Department of Physics \& Atmospheric Science, Dalhousie University,
Halifax, Nova Scotia B3H 3J5, Canada.\\
$^{2}$Centre for Computational Biology and Bioinformatics, School of Information Technology,
Jawaharlal Nehru University, New Delhi - 110 067, India.\\}
\ead{sengupta@mail.jnu.ac.in, andrew.rutenberg@dal.ca\hspace{1mm}}

\begin{abstract}
Ongoing sub-cellular oscillation of Min proteins is required to block minicelling in 
\textit{E. coli}. Experimentally, Min oscillations are seen in newly divided cells and 
no minicells are produced. In model Min systems many daughter cells do not oscillate 
following septation because of unequal partitioning of Min proteins between the daughter 
cells. Using the 3D model of Huang {\em et al.}, we investigate the septation process 
in detail to determine the cause of the asymmetric partitioning of Min proteins between 
daughter cells. We find that this partitioning problem arises at certain phases of the 
MinD and MinE oscillations with respect to septal closure and it persists independently of 
parameter variation. At most $85\%$ of the daughter cells exhibit Min oscillation following 
septation. Enhanced MinD binding at the static polar and dynamic septal regions, consistent with cardiolipin 
domains, does not substantially increase this fraction of oscillating daughters. We believe 
that this problem will be shared among all existing Min models and discuss possible biological 
mechanisms that may minimize partitioning errors of Min proteins following septation. 
\end{abstract}

\pacs{87.16.Ac, 87.17.Ee, 05.40.-a}
\submitto{\it Phys. Biol.\/}
\noindent{\it Keywords\/}: septation, \textit{Escherichia coli}, MinD, MinE, 
protein partitioning, oscillation, spatiotemporal pattern, subcellular 
localization, reaction diffusion, modeling. \\
Dated : \today \\

\maketitle

%%%%%%%%%%%%%%%%%%%%%%%%%%%%%%%%%%%%%%%%%%%%%%%%%%%%%%%%%%%%%%%%%%%%%%%%%%%%%%%%%%
\section{Introduction}

Cell division in \textit{Escherichia coli} is initiated by the formation 
of a ring of the protein FtsZ on the bacterial inner membrane. This FtsZ ring 
shrinks \cite{burdett} as the growing septum restricts the cytoplasmic channel 
connecting the two daughter cells. FtsZ ring formation is targeted to the 
mid-cell by two independent processes. Nucleoid occlusion prevents FtsZ ring 
formation over the nucleoids \cite{woldringh1,yu,bernhardt}, while polar 
FtsZ ring formation is prevented due to the oscillatory dynamics of the Min 
family of proteins. The pole-to-pole oscillation of MinD and 
MinE \cite{deboer,hale,hu} targets MinC to the polar inner membrane 
where it inhibits polar FtsZ ring formation \cite{cao,mukherjee} 
and prevents minicelling. 

Several deterministic \cite{meinhardt,hrdv,kruse1,kruse2,wingreen} 
and stochastic models \cite{howard,kerr,pavin,drew,tostevin} 
have been developed to explain the pole-to-pole oscillation pattern 
of the Min proteins. All these quantitative models have recovered 
oscillatory behavior, though they differ in their detailed interactions.

The FtsZ ring is the first element of the divisome to localize \cite{beckwith}. 
Induced disassembly of the FtsZ ring
can occur within a minute \cite{addinall}, and subsequent relocalization occurs
within minutes.  FtsZ can localize around potential division sites of daughter cells even before 
septation is complete \cite{margolin}. 
Min oscillations must persist or be quickly regenerated 
after septation to ensure that polar FtsZ ring formation is blocked 
in newly formed daughters. 

The experimental phenomenology of Min dynamics during septation has not yet been well 
characterized.  Early experiments \cite{hu,raskin} indicate that Min oscillations are 
qualitatively unaffected by partially constricted cells. 
Significantly, minicelling rates in wild-type {\it E. coli} cells 
are insignificant \cite{frazer}, and no non-oscillating daughter cells have been 
reported. These observations suggest that Min
oscillations persist or regenerate quickly in all daughter cells and, as a result,
block FtsZ ring formation at the poles of newly formed daughter cells.

In a pioneering study, Tostevin and Howard \cite{tostevin} addressed Min oscillations 
after cell division with a $1d$ stochastic model. Their model 
exhibited significant asymmetry in the distribution of Min proteins between the 
two daughter cells after division.  Approximately $20\%$ of their daughter 
cells did not oscillate due to such partitioning errors.  While systematic studies 
of partitioning errors have not been done, large asymmetries of concentrations 
between daughter cells have not been reported.  Tostevin and Howard suggested that rapid 
regeneration of Min proteins could quickly recover oscillations in non-oscillating 
daughters. However, no such cell-cycle dependent signal is seen in translation \cite{lut1} or, for
the {\em min} operon, in transcription \cite{arends}.  Moreover,  Min oscillations continue 
even when protein synthesis is stopped by chloramphenicol \cite{deboer}. This indicates that proteolysis 
rates are small, so that fast unregulated turnover of Min proteins (independent of the cell-cycle) is 
also not expected. 

In model systems, the MinD::MinE densities must be above a 
threshold or ``stability boundary'' for stable oscillations to be observed \cite{wingreen}.  
Experimentally, the position of the stability boundary is not precisely known though 
large overexpression of MinE does lead to minicelling \cite{pichoff,zhao}. 
We have explored how the distance of the parent cell from the stability boundary affects the
partitioning and hence the percentage of daughter cells that oscillate. 
While {\em in vivo} quantification of Min concentration has been done \cite{shih2002},
we have primarily varied the distance from the stability boundary by varying the concentration
of Min proteins in the parent cell within a reasonable range. This has the advantage of keeping the
stability boundary fixed. For completeness, we have also varied the model interaction parameters.
These are generally under-determined by experiment 
--- though diffusivities have now been measured {\em in vivo} \cite{meacci2006}.

In addition to varying existing parameters, we have also explored 
heterogeneous interactions along the bacterial length -- following \cite{drew}.  
The different phospholipids that comprise the {\em  E. coli} inner-membrane 
exhibit variable affinity for MinD \cite{dowhan1}.  Cardiolipin (CL) is preferentially
localized to polar and septal membranes in {\em E. coli} \cite{dowhan2,nanninga}.
The differences in MinD affinity for anionic phospholipids like CL implies enhanced 
MinD binding to the poles and the growing septum. We explore the implications of this 
midcell and polar enhancement on Min partitioning after septation. 

We study septation within the context of the model by Huang {\em et al.} \cite{wingreen}, 
which is a deterministic, $3D$ model without explicit MinD polymerization. This model is 
significantly different from the stochastic, $1D$, polymerizing model of Tostevin and 
Howard \cite{tostevin}.  Strikingly, we find that our partitioning errors are comparable 
to those seen by Tostevin and Howard \cite{tostevin} despite the differences in the models.
The model of Huang {\em et al.} still appears to 
be the best current model at recovering the Min oscillation phenomenology, though MinD 
polymerization appears to be called for experimentally \cite{rothfield} and has been used 
in several quantitative models of Min oscillation \cite{pavin,drew,tostevin}. Our aim is to 
understand how asymmetric partitioning result from the dynamics of Min oscillations and 
explore possible ways of achieving adequate partitioning of Min between daughter cells.
We analyze the origins of the partitioning error, and speculate about plausible partitioning 
mechanisms for Min proteins during the septation of {\em E. coli}. 

%%%%%%%%%%%%%%%%%%%%%%%%%%%%%%%%%%%%%%%%%%%%%%%%%%%%%%%%%%%%%%%%%%%%%%%%%%%%%%%%%%%
\section{Cell Division Model}
\label{model}

The model developed by Huang {\it et al.} \cite{wingreen} includes 
many of the interactions observed experimentally \cite{hu2,lutkenhaus,raychaudhuri,lackner}:
\begin{eqnarray}
\frac{\partial \rho_{D:ADP}}{\partial t} &= D_D\nabla^2\rho_{D:ADP}& - 
	\sigma_D^{ADP \rightarrow ATP}\rho_{D:ADP} + \delta_{mem}\sigma_{de}\rho_{de},   
	\label{DADP}\\
\frac{\partial \rho_{D:ATP}}{\partial t} &= D_D\nabla^2\rho_{D:ATP}& + 
	\sigma_D^{ADP \rightarrow ATP}\rho_{D:ADP} \nonumber \\ 
	&& - \hspace{2 mm} \delta_{mem}[\sigma_D + \sigma_{dD}(\rho_{d} + \rho_{de})]\rho_{D:ATP},   
	\label{DATP} \\
\frac{\partial \rho_{E}}{\partial t} &= D_E\nabla^2\rho_{E}& + 
	\delta_{mem}\sigma_{de}\rho_{de} - \delta_{mem}\sigma_{E}\rho_{d}\rho_E,  
	\label{E} \\
\frac{\partial \rho_{d}}{\partial t} &=& -\sigma_E\rho_d\rho_E(M) +  
	[\sigma_{D} + \sigma_{dD}(\rho_d + \rho_{de})]\rho_{d}\rho_E  
	\label{d}\\
\frac{\partial \rho_{de}}{\partial t} &=& -\sigma_{de}\rho_{de} +  \sigma_{E}\rho_{d}\rho_{E}(M),  
	\label{de}
\end{eqnarray}
where $\rho_{DADP}$, $\rho_{DATP}$ and $\rho_{E}$, are the 
cytoplasmic densities of MinD:ADP, MinD:ATP and MinE respectively 
and $\rho_{d}$, $\rho_{de}$ are the densities of membrane-bound MinD 
and MinDE complex, respectively.  The rates of 
binding of MinD:ATP to the bare membrane, the cooperative binding of 
MinD:ATP to membrane bound MinD:ATP, the binding of cytoplasmic MinE to
membrane bound MinD:ATP, and the hydrolysis rate of MinD:ATP from the
membrane under activation by MinE are given by 
$\sigma_{D}$, $\sigma_{dD}$, $\sigma_{E}$, and $\sigma_{de}$, respectively.
The bacterium was modeled as a cylinder of length $L$ and
radius $R$, with longitudinal interval $dx=0.0521$ and radial 
interval $dr=0.0416$, and with poles represented by flat, circular end-caps. 
Lateral growth is significantly reduced during septation \cite{wold2,wientjes}, so we
accordingly keep $L$ constant.  The density of cytoplasmic MinE at the membrane surface is 
$\rho_{E}(M)$, while 
$\delta_{mem} \equiv \delta(r-R) + \delta(z) + \delta(z-z_{L}) + \delta(z-L/2)\theta(r-r_s(t))$ 
limits reactions to the bacterial inner membrane.  The last
term denotes the growing septum at mid-cell, with 
$r_s(t) \in [0,R]$ being the radius of the circular open portion of the 
cylindrical cross-section at mid-cell. 
Diffusion is not allowed across the septum (for $r>r_s$), while 
membrane interactions take place independently on either side of the growing septum.

Pre-septation Min oscillations 
were allowed to stabilize in a cell with length $L=5 \mu m$ before the process 
of septation was initiated. Septation was initiated at $10$ or more uniformly distributed 
phases of the Min oscillation period to determine the effect of this phase on 
the partitioning of Min proteins. Since the detailed septal closure dynamics of {\it E. coli} 
are not well constrained experimentally, we assume linear inward growth of the septum with the 
midcell septal radius given by 
\begin{equation}
	r_s(t) = R\left(1 - (t-t_s)/t_r \right), \ \ \ \ t> t_s
\end{equation}
where $t_r$ is the duration of septation 
and $t_s$ is the time at which septation starts. The area of the growing
septum is then $A(t)=\pi(R^2 - r_s^2)$ for $t \in [t_s,t_s+t_r]$ ($A=0$ for $t<t_s$ while
$A=\pi R^2$ for $t>t_s+t_r$). 
This process of septal closure mimics the process of septal growth discussed by 
Burdett and Murray \cite{burdett}. 
Since MinD:ATP has a greater affinity for 
anionic phospholipids such as CL \cite{dowhan1,dowhan3} and since CL domains 
are found to be localized around the cell poles and septal regions
\cite{dowhan2,nanninga}, we also considered the case in which the rate of attachment of 
MinD:ATP ($\sigma_D$) was enhanced at the polar and septal membranes (by an amount $\sigma_{Dmax}$) compared to 
the attachment rate elsewhere on the curved surface of the cylindrical cell ($\sigma_{D0}$). 

\begin{figure} 
\centering
\includegraphics[width=4.0cm]{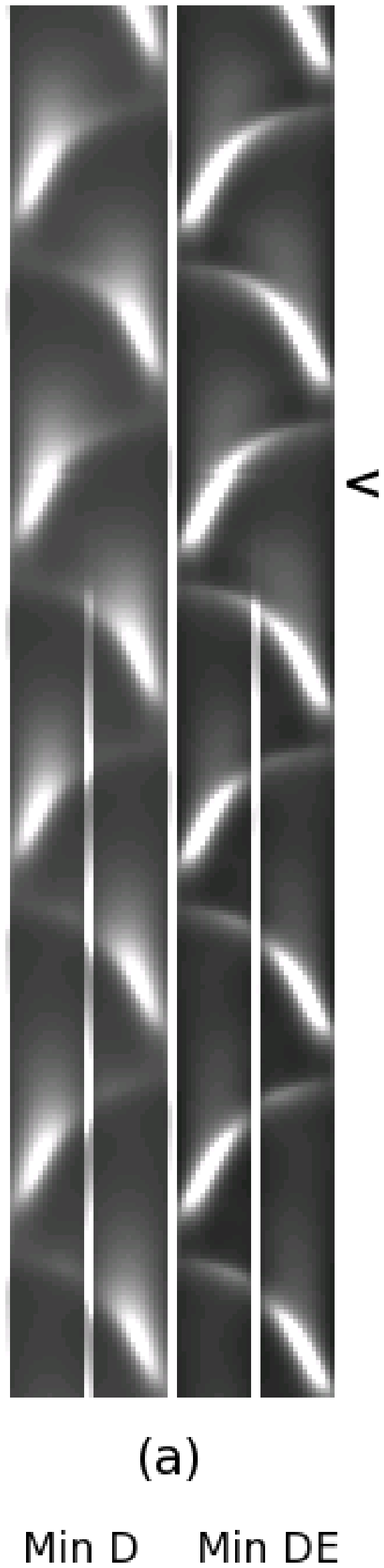} 
\includegraphics[width=4.0cm]{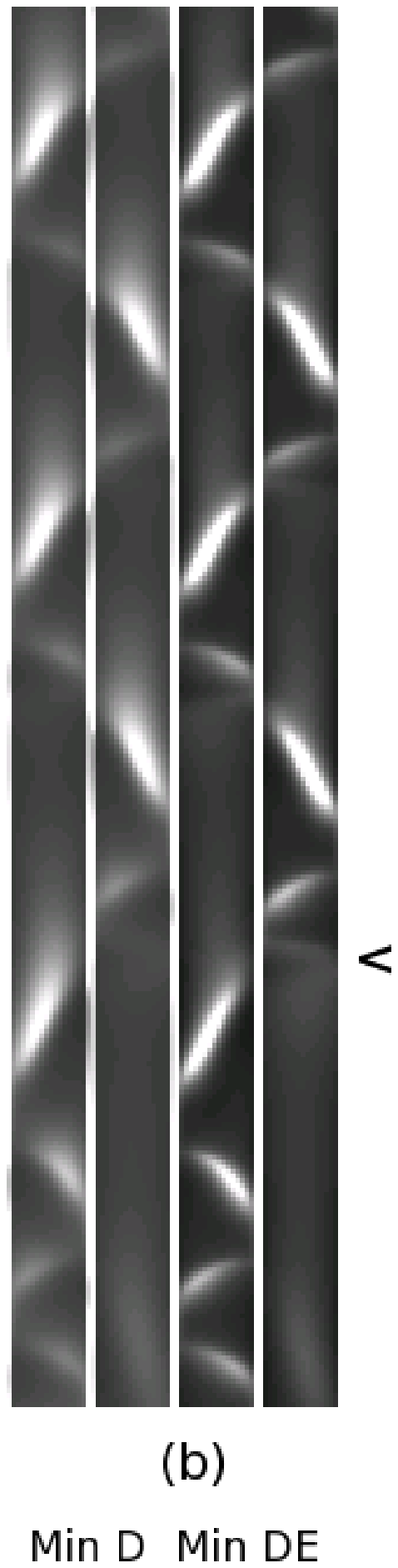} 
\includegraphics[width=4.0cm]{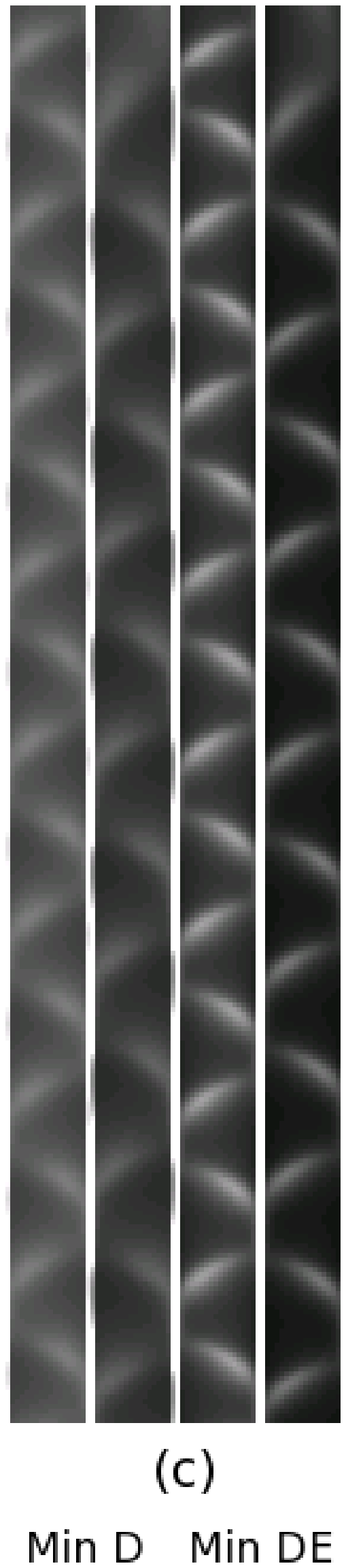} 
\caption{The time-development of membrane-bound MinD and MinDE during and after
septation, represented as space-time plots.
White and black indicate high and low linear densities,
respectively. Time increases from top to bottom (total duration of 
$300$ s is shown) while the bacterial length runs from left to right ($L=5 \mu m$)
for each of MinD and MinDE. 
(a) Oscillations in parent cell starting from $100$ s before and ending $200$ s after septation. 
Membrane-bound MinD and MinDE are shown in the first and second columns respectively, as indicated.
The arrowhead marks the beginning of the septation process and emerging white bar at midcell 
corresponds to the growing septum. (b) Oscillations just before and after the end of septation.
The arrowhead marks the end of the septation process and formation of two independent daughter 
cells. Oscillations continue in the left daughter cell through septation but are disrupted and 
then regenerated in the right daughter cell after septation is complete. (c) Oscillations in 
both daughters after completion of septation. A significant asymmetry of Min partitioning between 
the two daughter cells is apparent. The parameters used in this figure are 
$\rho_D=1150 \mu m^{-1}$, $\rho_E=350 \mu m^{-1}$, 
$D_D=D_E=2.5 \mu m^{2}/sec$, 
$\sigma_{D0}=0.025 \mu m/sec$, $\sigma_{dD}=0.0015 \mu m^{3}/sec$, 
$\sigma_{de}=0.7/sec$, $\sigma_E=0.093 \mu m^{3}/sec$,
$\sigma_{Dmax}=0.1 \mu m/sec$. 
The three subfigures are contiguous in time.}
\label{sep} 
\end{figure}

Fig.~\ref{sep}(a) and 1(b) show oscillations in the parent cell during the process 
of septation while Fig.~\ref{sep}(c) shows oscillations in both daughters 
after septation. A septation duration $t_r=512$ seconds was chosen to be 
consistent with the proportion of septating cells observed in culture \cite{woldringh}.
Significantly faster septation ($t_r=350$ s) does not affect our results.

%%%%%%%%%%%%%%%% FIGURE 1 %%%%%%%%%%%%%%%%%%%

%%%%%%%%%%%%%%%%%%%%%%%%%%%%%%%%%%%%%%%%%%%%%%%%%%%%%%%%%%%%%%%%%%%%%%%%%%%%%%%
\section{Results}

%%%%%%%%%%%%%%%%%%%%%%%%%%%%%%%%%%%
\subsection{Varying Min concentration}

We examined the effect of varying the MinD and MinE densities in the parent cell on the 
partitioning of Min between daughter cells. For this purpose, we generated $420$ 
sets of different parent cell densities ($\rho_D,\rho_E$).  For each set, the 
initiation time of septation $t_s$ was varied uniformly over the oscillation period $T$ of the
parent cell with at least $10$ different phases sampled for each parent cell.
Min partitioning information was noted at the end of the septation, and the simulation was run
for more than 15 minutes after the end of septation to 
see whether Min oscillations were regenerated in the daughter cells. 
 
%%%%%%%FIGURE 2 %%%%%%%%%%%%%%%%%%%%%%%%%%%%
\begin{figure} 
%\centering
%\includegraphics[angle=270,width=12cm]{noenhance-jan22-modified.ps} 
\includegraphics[angle=270,width=16cm]{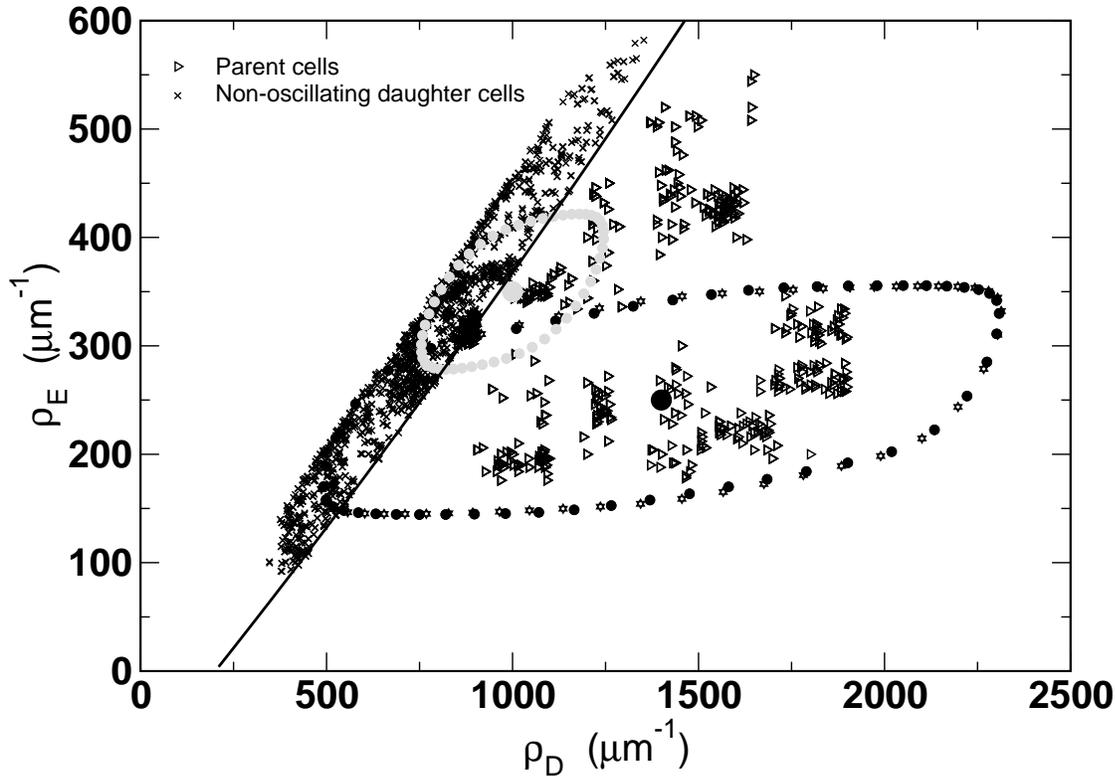} 
\caption{Scatter plot of linear MinD and MinE densities in the parent cell 
(open triangles) and non-oscillating daughter cells ($\times$). Indicated
by the solid line is the approximate stability curve for $L= 5 \mu m$ cells, 
separating oscillating and non-oscillating daughter cells. The large black and grey 
filled circles indicate example parent cells that lead to $84\%$ and $58\%$ oscillating 
daughter cells respectively. The smaller black and grey filled circles denote the 
corresponding linear densities of daughter cells produced after cell division. The
open stars correspond to the black filled circle but with a 
septation duration of $t_r=350$ seconds, all other points correspond to $t_r=512$ seconds.
Parameters are as specified in Fig.~1, but with $\sigma_{Dmax}=0$.}
\label{noen} 
\end{figure}
%%%%%%%%%%%%%%%%%%%%%%%%%%%%%%%%%%%%%%%%%%%%%%%%%%%%%

Fig.~\ref{noen} shows the linear density of MinD and MinE in parent cells (open triangles).
Daughter cells with a variety of phases of septation start times are shown (smaller black and grey 
filled circles) for two representative parent cells (larger black and grey filled circles) close to 
and far from the stability boundary (approximately indicated by the black line), 
respectively.   For oscillations to restart or continue in daughter cells, the 
the ratio of MinD:MinE must be greater than $\approx 2.7$.
Inadequate partitioning of Min results in daughter cells ($\times$)
having Min densities which fall below this threshold.  The partitioning for the pole-to-pole oscillating 
MinD is worse than for the more midcell MinE, resulting in an asymmetric donut-shaped distribution of 
daughter cell densities for a given parent cell. Since the MinE ring closely follows the MinD cap there 
is a correlation between the MinD and MinE partitioning --- extending the asymmetric donuts along the 
diagonal. For parent cell densities close to the oscillation threshold, a large fraction of daughter cells 
do not a oscillate. Away from the threshold a smaller fraction do not oscillate.  Varying the duration of 
septation by moderate amounts does not change the partitioning, as illustrated by the nearly identical 
donuts for $t_r=350s$ (open stars) and $t_r=512s$ (black circles).

%%% FIGURE 3 %%%%%%%%%%%%%%%%%%%%%%%%%%%%%%%%%%%%%%%%
\begin{figure} 
%\centering 
%\includegraphics[angle=270,width=7cm]{noenhance-fraction-jan22.ps} 
%\includegraphics[angle=270, width=7cm]{scaled-dstn-noenhance-new.ps} 
%\hspace{-0mm}
\includegraphics[angle=270, width=10cm]{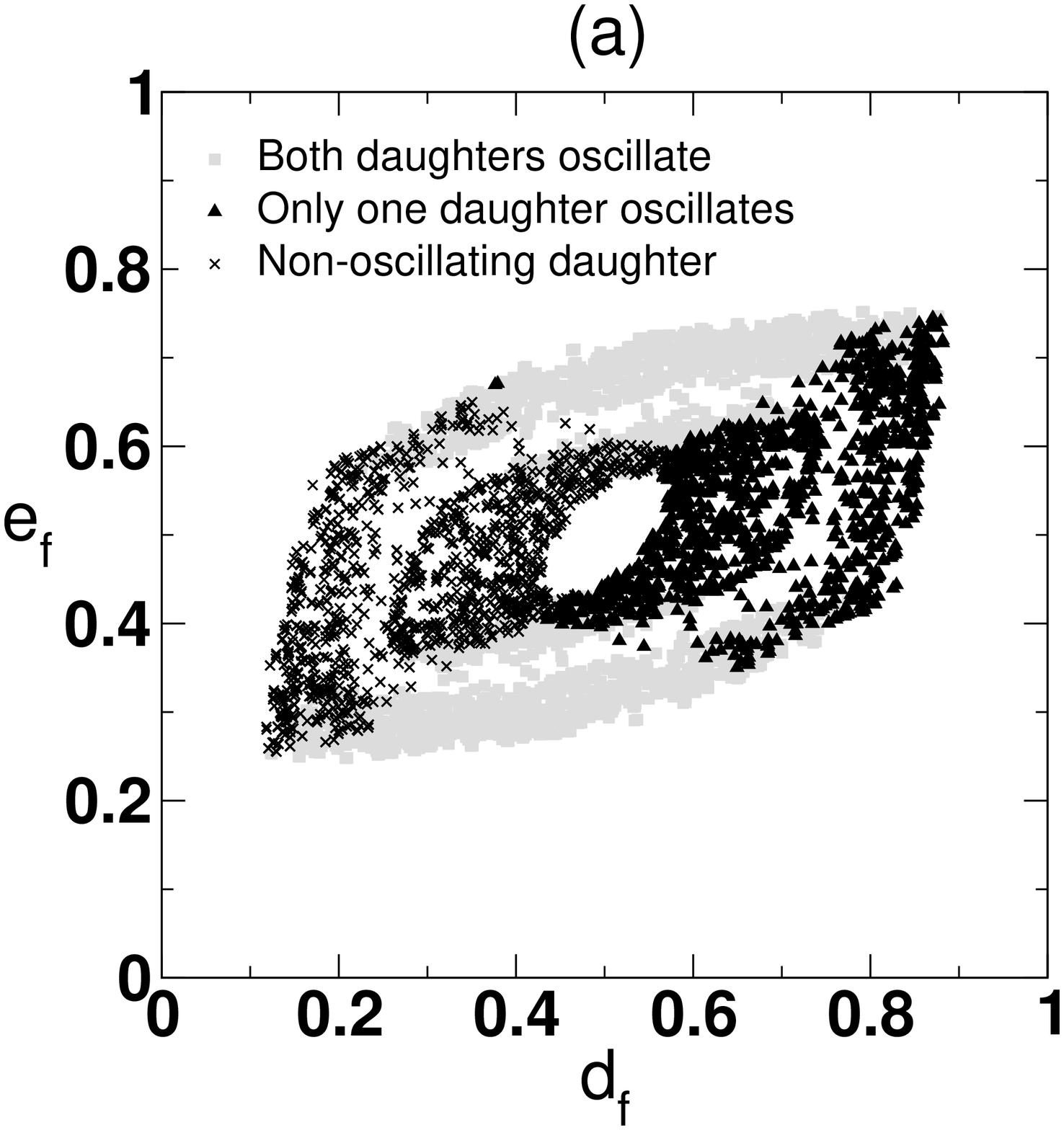} 
\hspace{-24mm}
\includegraphics[angle=270,width=10cm]{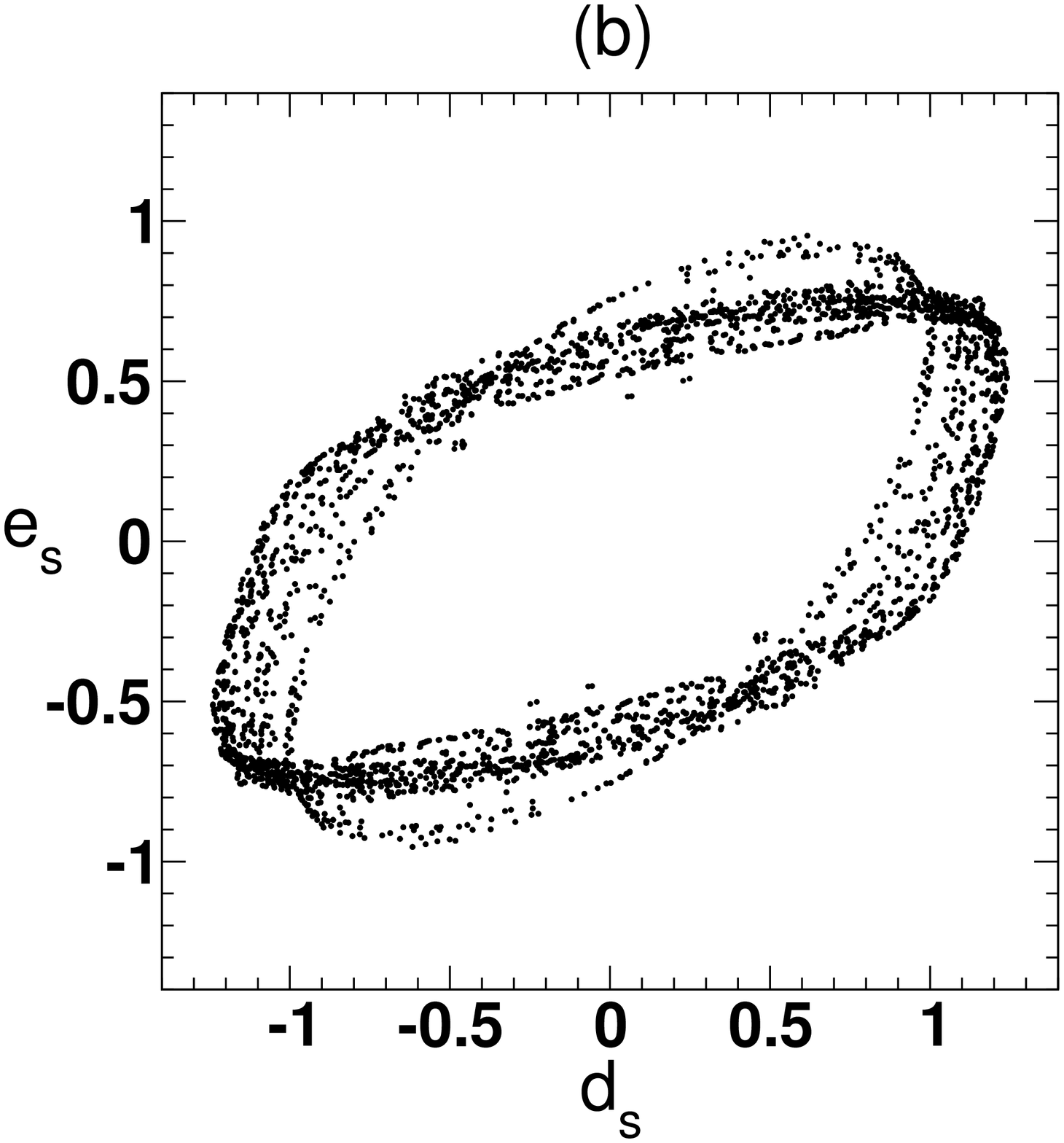} 
\caption{(a) Fractions, $d_f$ vs. $e_f$ of MinD 
and MinE, respectively, in oscillating as well as nonoscillating daughters 
for all parent cells. Black $\times$ indicates non-oscillating daughter cells, 
while grey filled circles indicate oscillating daughter cells in cases where 
both daughter oscillates. Filled upper triangles correspond to fractions in 
the oscillating daughter for cases where only one daughter oscillates. The 
two daughter cells of a given parent are symmetrically placed around 
$d_f=e_f=0.5$. (b) A plot of the scaled relative fractions of MinD vs. MinE
in the two daughter cells.}
\label{scal1} 
\end{figure}
%%%%%%%%%%%%%%%%%%%%%%%%%%%%%%%%%%%%%%%%%%%%%%%%%%%%%

In Fig.~\ref{scal1}(a) we show all of the partitioning donuts on one plot, where $d_f$ and $e_f$ are 
the fraction of MinD and MinE in the two daughter cells, respectively.  The absence of any 
daughter cells in the central region, near $d_f=e_f=0.5$, 
shows that simultaneous equipartitioning of both MinD and MinE is never observed.  
While there is always a septation start-time $t_s$, relative to the parent cell oscillation, 
that leads to perfect partitioning of MinD {\em or} MinE, there is no phase that leads to 
perfect partitioning of {\em both} MinD and MinE.  This ``donut hole'' 
is a manifestation of the phase lag between MinD and MinE 
oscillations, i.e. the timing of maximal MinD at midcell is ahead of the timing of maximal MinE.  
To make this clear,  in Fig.~\ref{scal1}(b) we have scaled all of the partitioning donuts by their RMS radius, 
$r_{av} \equiv \sqrt{\langle (d_f-0.5)^2 + (e_f-0.5)^2\rangle}$, 
and plotted the scaled densities $d_s \equiv (d_f-0.5)/r_{av}$ vs. 
$e_s \equiv (e_f-0.5)/r_{av}$.  Relative to $r_{av}$, there are no 
phases that approach symmetric partitioning of both MinD and MinE. 

\begin{figure} 
%\centering 
%\includegraphics[angle=270, width=12cm]{avg-pp-noenhance-new.ps} 
\includegraphics[angle=270, width=16cm]{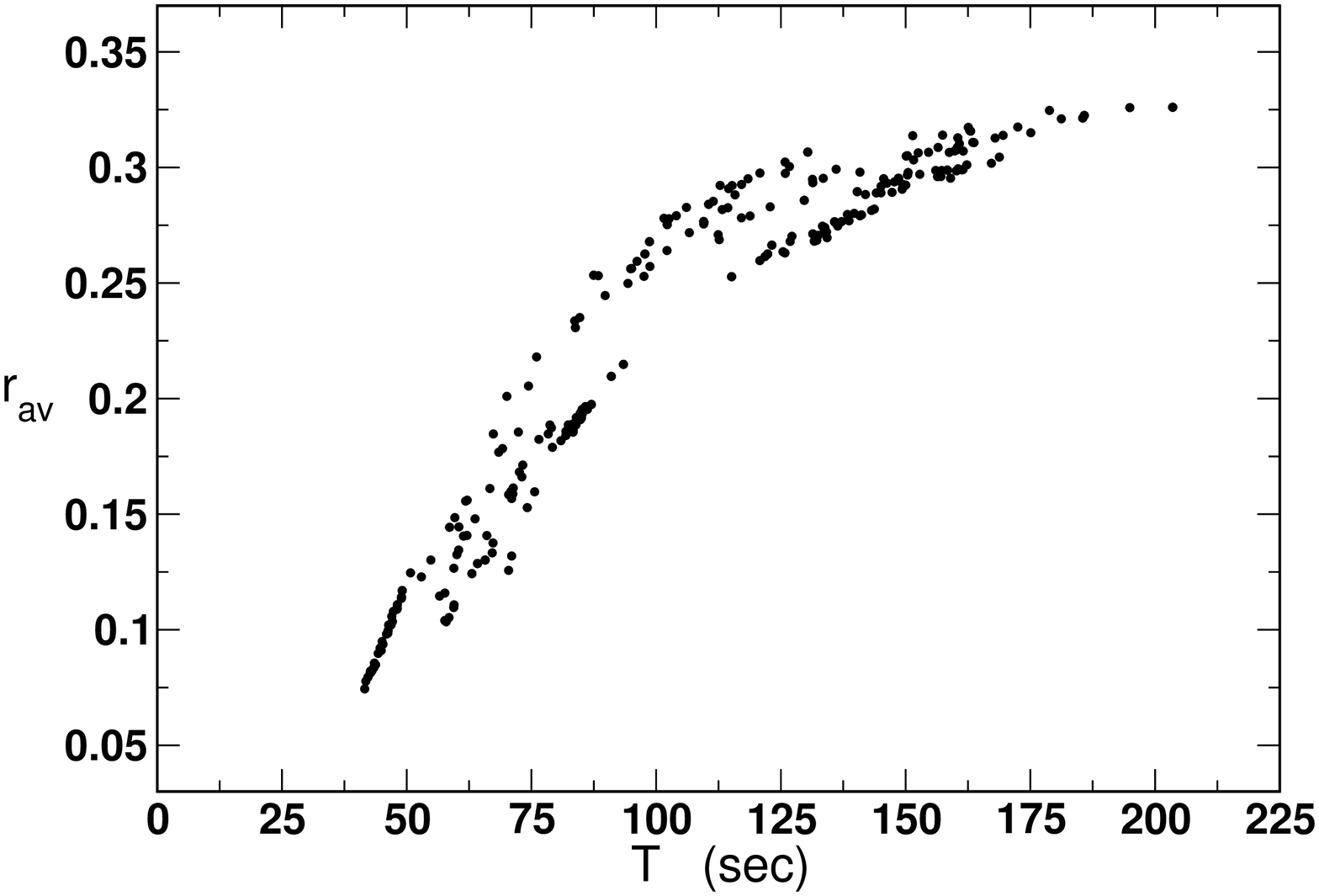} 
\caption{The RMS magnitude of partitioning error $r_{av}$ vs. the oscillation period of the mother cell $T$. The period
increases as the mother cell Min concentrations are moved away from the stability boundary shown in Fig.~\protect\ref{noen}. 
While there is no precise scaling collapse, the trend is for less accurate partitioning as distance from the stability boundary
(and hence $T$) increases --- maintaining the non-oscillating daughters shown in Fig.~\protect\ref{noen}.  
We do not find a significant dependence of $r_{av}$ on the septation duration $t_r$.} 
\label{scal2} 
\end{figure}

We also plotted $r_{av}$ against the oscillation period $T$ of the parent cell in 
Fig.~\ref{scal2} to determine if the RMS radius scales with the period of oscillation
of the parent cell. We do not see perfect collapse but $r_{av}$ increases with period 
away from the stability boundary, indicating that the two partitioning donuts (formed 
by the small black or grey filled circles) shown in Fig.~\ref{noen} are representative. 

%%%%%% FIGURE 4 %%%%%%%%%%%%%%%%%%%%%%%%%%%%%%%%%%%%%
%%%%%%%%%%%%%%%%%%%%%%%%%%%%%%%%%%%%%%%%%%%%%%%%%%%%%

%%%%%%%%%%%%%%%%%%%%%%%%%%%%%%%%%%%%%%%%%%%%%%%%%%%%%%%%%%%%%%%%
\subsection{Enhanced MinD binding at poles and septum}

To see whether a distinct phospholipid composition of the closing septum could affect 
the partitioning, we enhanced MinD:ATP binding ($\sigma_D$) at the cell poles and the 
growing mid-cell septum.  
The degree of enhancement was constrained by the practical 
requirement that it did not disrupt steady oscillations in the parent cell before $t_s$. 
This restricted the polar enhancement $\sigma_{Dmax}$ to less than ten times the base 
value of $\sigma_{D0}=0.025 \mu m/sec$. This is consistent with the affinity of MinD:ATP 
for anionic phospholipids like cardiolipin, which is nine times higher than its affinity 
for zwitterionic phospholipids \cite{dowhan2}. 
The enhancement of MinD:ATP binding at 
the poles and septum slightly increased the oscillation period in the parent cell 
by increasing the time for dissociation of membrane-bound MinD:ATP by MinE. 

To analyze the effect of enhanced MinD binding at the poles and growing septum on 
the number of oscillating daughters, we compared the results from  50 parameter sets 
with and without septal and polar enhancement. 
In this comparison, the concentrations 
of MinD and MinE were varied while all other parameters were 
kept fixed and $\sigma_{Dmax}=0.1 \mu m/sec$ or $\sigma_{Dmax}=\sigma_{D0}$. 
The overall percentage of oscillating daughter cells increased by a small 
amount (~2\%) when enhanced polar and septal MinD:ATP attachment rates were used.
More specifically, for parent cell density close to the stability 
threshold (large grey filled circle in Fig.~\ref{noen}), the enhancement of MinD:ATP 
binding at the poles and septum led to a modest increase (at most $2\%$) in the number 
of daughters which restart oscillations after septation. However, for parent cell densities 
far from the stability threshold (large black filled circle in Fig.~\ref{noen}) no 
significant increase in the number of oscillating daughters was obtained with enhanced 
MinD:ATP binding at the poles and growing septum.  

%%%%%%%%%%%%%%%%%%%%%%%%%%%%%%%%%%%%%%%%%%%%%%%%%%%%%%%%%%%%%%%%
\subsection{Varying interaction parameters}

In another attempt to increase the fraction of oscillating daughter cells 
after septation, we explored the parameter space of interactions in the Huang 
{\it et al.} model \cite{wingreen}. Since most of the parameters are 
experimentally under-determined, some flexibility is possible in the choice 
of parameters while insisting upon stable oscillations. In this context, 
the Min concentration, diffusivities, reaction rates, and $\sigma_{Dmax}$
were all independently varied over plausible ranges for a fixed cell length. The parameter 
space was explored to move towards symmetric partitioning of MinD and MinE 
in non-oscillating daughter cells. Each parameter was varied over a range 
spanning almost an order of magnitude relative to the benchmark values which 
were chosen to be the parameters specified in Huang {\it et al.} However, no 
improvement upon the best $85\%$ figure 
(obtained with or without polar and septal enhancement of the MinD binding rate) was obtained. 

%%%%%%%%%%%%%%%%%%%%%%%%%%%%%%%%%%%%%%%%%%%%%%%%%%%%%%%%%%%%%%
\subsection{Phase dependence of partitioning}
\begin{figure} 
%\centering 
%\includegraphics[angle=270, width=7cm]{de-peaks-b0-osc.ps} 
%\includegraphics[angle=270, width=7cm]{de-peaks-b0-nonosc.ps} 
\includegraphics[angle=270, width=16cm]{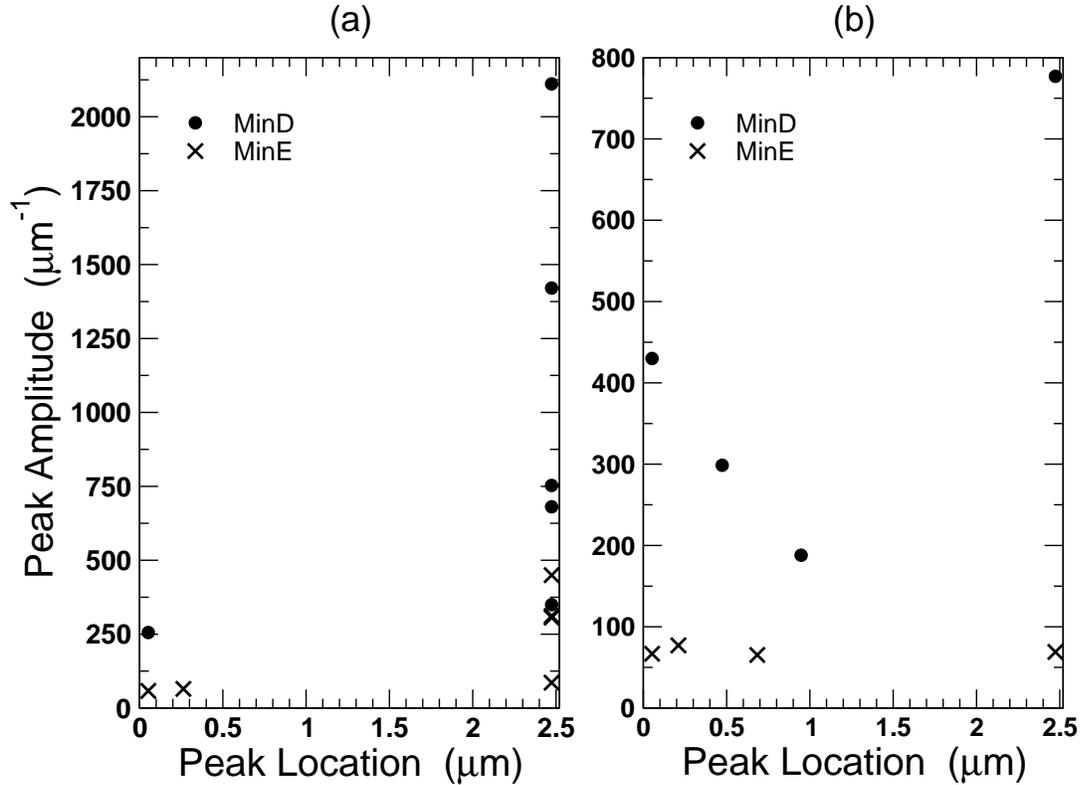} 
\caption{Non-polar ``lateral'' peak location of radially integrated MinD and MinE 
at the end of septation for phases which lead to (a) two oscillating daughters and 
(b) one oscillating daughter. Oscillations are observed in both daughters for 12 out
of the 20 septating phases. Only half the cell length is plotted since the peak 
locations for different septating phases are symmetric about the mid-cell. In general, 
two-oscillating daughters result from a strong central peak of MinD at the completion 
of septation while one-oscillating daughter results from polar peaks, and hence weak 
and non-central lateral peaks. Parameters are as specified in Fig.~1, except for 
$\rho_D=1400 \mu m^{-1}$, $\rho_E=250 \mu m^{-1}$, $t_r=512$ seconds, 
and $\sigma_{Dmax}=0$.}
\label{phase} 
\end{figure}

Why do we never see $100\%$ of the daughter cells oscillating? The pattern of end-to-end 
oscillation of MinD continues largely unchanged throughout septation (see, e.g., 
Fig.~\ref{sep}), even as the period lengthens somewhat, so that when the {\em closure} 
of the septum coincides with MinD being localized predominately at one pole then the 
MinD will be badly partitioned between the two daughter cells. In Fig.~\ref{phase} we 
plot the longitudinal position of the radially integrated MinD and MinE peaks away 
from the cell poles at the end of septation when $t=t_s+t_r$.  Fig.~\ref{phase}(a) 
shows parent cells that lead to two oscillating daughters, while Fig.~\ref{phase}(b) 
shows parent cells that lead to only one oscillating daughter. We see that two oscillating 
daughters typically result from septation events where both MinD and MinE have a 
substantial peak at the mid-cell. 
When two oscillating daughters result despite polar maxima of MinD and MinE, a substantial midcell
accumulation of MinD is also present.  A non-oscillating daughter cell is typically produced 
when MinD has a large peak near one pole. 

%%%%%% FIGURE 5 %%%%%%%%%%%%%%%%%%%%%%%%%%%%%%%%%%%%%
%%%%%%%%%%%%%%%%%%%%%%%%%%%%%%%%%%%%%%%%%%%%%%%%%%%%%%%%%%%%%%%%%%%%%%%%%%%%%% 

\begin{figure} 
%\centering 
%\includegraphics[angle=270, width=14cm]{de-profiles-b0-cuts6-1-3.ps} 
\includegraphics[angle=270, width=16cm]{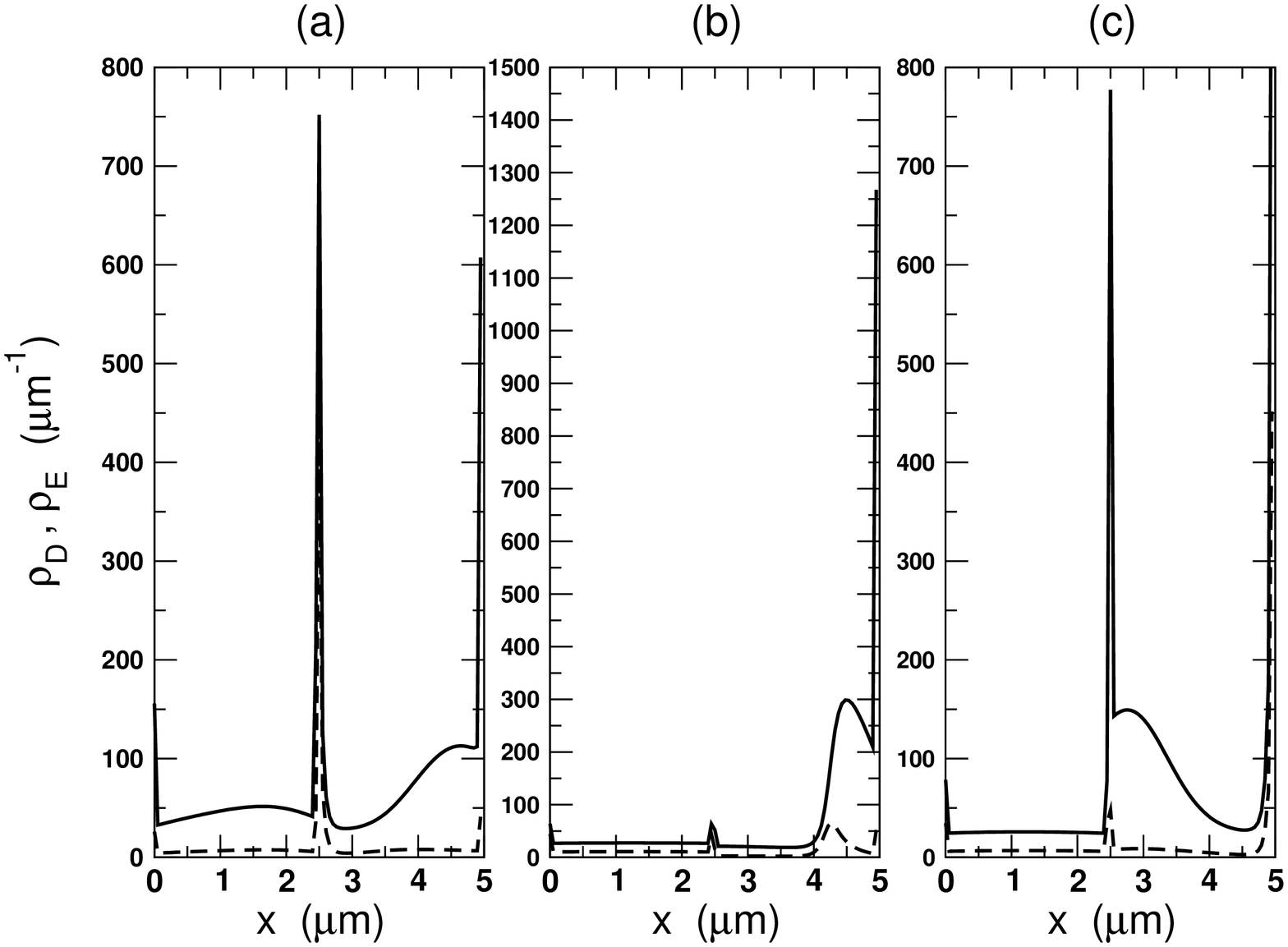} 
\caption{Profile along the cell length $x$ of radially integrated (linear) densities of MinD (solid line) and 
MinE (dashed line) for three different phases of septation, at the time of septal closure. (a) leads to two oscillating 
daughter cells and exhibits strong central MinD and MinE peaks, (b) leads to only one oscillating daughter cell and exhibits a
strong polar peak of both MinD and MinE, and (c) leads to only one oscillating daughter cell and exhibits a strong polar peak 
of MinE. The parameters used are the same as in the previous figure.}
\label{prfl} 
\end{figure}

Fig.~\ref{prfl} illustrates the spatial profile of radially integrated MinD and MinE 
for three different phases at the end of septation. Fig.~\ref{prfl}(a) corresponds 
to a phase where oscillation restarts in both daughters after septation. Adequate 
partitioning is reflected in large peaks of radially integrated MinD (solid line) 
and MinE (dashed line) near the midpoint of the cell.  Fig.~\ref{prfl}(b) corresponds 
to a phase where inadequate partitioning is manifest in the large peaks of radially 
integrated MinD (solid line) and MinE (dashed line) near one pole of the cell. Only 
one oscillating daughter results. Fig.~\ref{prfl}(c) shows a peak in the radially 
integrated MinD and MinE near the midcell and pole respectively. The resulting 
inadequate partitioning of MinE between the two daughters ensures that the ratio
of MinD:MinE falls below the threshold required to regenerate oscillations in 
one of the daughters. This leads to a non-oscillating daughter and corresponds to the points 
with a large midcell MinD peak in Fig.~\ref{phase}(b).  

%%%%%% FIGURE 6 %%%%%%%%%%%%%%%%%%%%%%%%%%%%%%%%%%%%%
%%%%%%%%%%%%%%%%%%%%%%%%%%%%%%%%%%%%%%%%%%%%%%%%%%%%%%%%%%%%%%%%%%%%%%%%%%%%%% 

\section{Discussion and Conclusions}

We have explored the impact of MinD and MinE concentration, interaction parameters, and 
end-cap and septal cardiolipin patches on the partitioning of Min proteins between
daughter cells after septation in the model of Huang {\em et al.} \cite{wingreen}. 
While concentration close to the 
stability threshold for oscillations led to less than $50\%$ of daughter cells 
oscillating after septation, no combination of concentration, interaction parameters,
and/or cardiolipin patches led to more than $\approx85\%$ of daughter cells oscillating
after septation. These results are comparable to those of Tostevin and Howard \cite{tostevin},
despite significant differences in the Min models that were used. They studied a stochastic one-dimensional
model with explicit MinD polymerization, while we used a deterministic three-dimensional model without
filamentous MinD structures.  
We do not expect that the inclusion of stochastic effects would significantly change our results,
following \cite{kerr}.

We found that plotting the MinD vs. MinE densities in the daughter cells leads to a donut 
structure around the parent cell densities, and that varying the phase of the septal 
closure with respect to the end-to-end Min oscillation of the parent cell leads to daughter 
Min densities varying around the donut. The ``missing hole'' of the donut, i.e. the absence 
of daughter cells with the same Min densities as the parent cell, arises from the 
phase-difference between the leading MinD cap-forming and lagging MinE ring-forming 
oscillations. Furthermore, we find that there is always a phase of septation timing that 
leads to non-oscillating daughters.  We believe that this is a fundamental aspect of 
end-to-end Min oscillation: when the MinD cap is at one pole, the distal pole is stable.  
This should be a generic feature of all Min oscillation models. The robustness of the 
best percentage of oscillating daughters under changes in concentration, parameter 
variation, heterogeneous perturbations, model variation, dimensionality, and stochastic 
effects support this conclusion. 

How might {\em E. coli} achieve its (observed) negligible level of minicelling? We see four basic
possibilities.

As suggested by Tostevin and Howard \cite{tostevin}, the non-oscillating daughters 
could be rescued by rapid regeneration of Min concentration. 
This would require Min synthesis to be regulated in a cell-cycle dependent
manner. Because the average concentration of the two daughter cells equals 
their parent cell, rapid synthesis leading to recovery in one daughter cell would 
lead to a spike in Min concentration right after septation. However, there is 
no evidence of such fine-tuned regulatory control, or cell-cycle dependence, 
of Min concentration \cite{lut1,arends}. Moreover, lack of adequate partitioning would give 
rise to substantial asymmetry of Min proteins in the two daughter cells 
that should be apparent in experimental studies --- especially
with the simple inducible promoters (not actively regulated) typically used in Min-GFP fusion 
studies \cite{deboer,hale,hu,raskin,shih2002,meacci2006,rothfield,hu2}. 
In our simulations we found that the fraction of the parent MinD and MinE in daughter cells can 
be as low $15\%$ and $25\%$ respectively. The lack of any reports
of such large visible asymmetries argues against rapid Min regeneration. 

The partitioning problem can be avoided if the Min oscillations 
``double-up'' before septation, leading to two symmetric oscillations
in the two halves of the parent cell. A closing septum would then maintain 
symmetric Min distributions in the daughter cells. Indeed, we were hoping 
to promote this effect with the introduction of cardiolipin patches at poles
and septum --- without success. While there has been one experimental report of a 
doubling of oscillation for deeply constricted cells \cite{hu}, this must be
approached with caution due to the difficulty of distinguishing partial from full
septation.  We never found any evidence for doubling up of 
oscillations in our simulations. In all cases, we found that oscillations 
continue until just before the end of septation.  Indeed, the Min oscillation 
wavelength of $\approx 8 \mu m$ seen in filamentous cells \cite{deboer}
would suggest that it is difficult to spontaneously generate $L=2 \mu m$
oscillations while significant connection between the two ends of the parent
cell remains.  

Distortion and/or disruption of the Min oscillation by the growing septum before septal closure 
might also lead to symmetric partitioning of Min between the daughter cells. We do find that 
MinD binding to the sides of the growing septum improves partitioning. 
This was evident by comparing the partitioning for a finite septation time 
($t_r=512$ seconds) with instantaneous septation ($t_r=0$). In the latter case, 
no MinD can accumulate on the septum before the daughter cells are separated. 
This resulted in highly skewed Min distributions between the two daughter cells 
(results not shown). However, significant partitioning errors still occur with 
gradual septal growth. Moreover, no significant improvement in partitioning was 
observed when the MinD binding was enhanced at the midcell. 
We also found that Min oscillation was often temporarily disrupted in one daughter cell 
despite acceptable partitioning for oscillation in both daughters. The time required 
for recovery of steady oscillations was sometimes as large as 15 minutes. 
This is much larger than the dynamical time-scale of FtsZ rings \cite{margolin}, 
though, as shown by Tostevin and Howard \cite{tostevin}, 
stochastic effects may eliminate or significantly decrease the regeneration time of oscillations.
Disruption of the Min oscillation in both daughter cells by the late stages of septation may therefore 
be a viable partitioning mechanism {\em in vivo} especially if the resulting uniform distribution of Min is 
sufficient to block septation \cite{deboer89} in the face of fast FtsZ 
dynamics \cite{margolin} while the Min oscillation is being regenerated. 
However, in our model we did not observe disruption in both daughter cells even with enhanced MinD binding
at the growing septum. 

Finally, the cell may coordinate the septal closure with the Min oscillation. As seen 
in Fig.~\ref{sep} there are a number of phases where {\em both} daughter cells oscillate
after septation.  As shown in Fig.~\ref{phase}(a), and illustrated in Fig.~\ref{prfl}(a), most of 
those phases correspond to midcell MinD and MinE peaks. Triggered septal closure that occurs 
only at these phases would always recover Min oscillation in both daughters. 
Such triggered septal closure could result from the participation of the C-terminal domain of MinC 
in FtsZ ring {\em disassembly} towards the end of of septation \cite{shiomi}. 
Since septation occurs in $\Delta min$ mutants \cite{deboer89}, any such effect would have to 
accelerate septation rather than cause it.  Narrow constrictions have been observed in cryoelectron 
tomography studies of {\em Caulobacter crescentus} \cite{judd}, though too infrequently to indicate a significant
septation pause.  In {\em E. coli}, mutations of the N-terminal domain of FtsK lead to the stalling of 
septation at a very late stage with deep constrictions \cite{begg}, 
leading to speculation about pores between the daughter cells before septal closure \cite{donachie}. The
triggered septal closure discussed here would only require a pause (or speed-up) of at most one half period of 
the Min oscillation that could be lifted (or imposed) by the MinC at midcell. 

The challenge lies in understanding how Min oscillations can persist or be regenerated in 
both daughter cells after septation, in the face of partitioning errors due to the 
end-to-end oscillation of the Min proteins.  Without one or more of 
the additional mechanisms discussed above, we expect significant partitioning
errors, leading to non-oscillating daughters, in all Min models. 
Experimental characterization of the Min oscillations during and after septation, 
and quantitative assessment of Min partitioning between the daughter cells will be 
invaluable in sorting out which of these four partitioning mechanisms, or what 
combination of these four mechanisms, plays a role in {\em E. coli}. We believe
that the last mechanism, of triggered septal closure, is most likely the dominant mechanism 
{\em in vivo}.  Reproducing Fig.~\ref{scal1} from experimental images of newly septated cells should
be straightforward if both MinD and MinE have distinct fluorescent tags (see, e.g. \cite{shih2002}).
The average of each fluorescent signal of the two daughter cells can be used to independently 
scale the corresponding MinD or MinE signal, without the need for calibration even in the 
face of photo-bleaching. Non-regenerating mechanisms of partitioning, such as septal triggering, 
would lead to a ``double-bar'' pattern of MinD vs MinE densities in the daughter cells 
(looking like ${- \atop -}$) rather than the connected donuts seen in Fig.~\ref{scal1}. 

\ack We thank Benjamin Downing and Manfred Jericho for useful discussions. This work was 
supported by the Canadian Institute of Health Research.

%%%%%%%%%%%%%%%%%%%%%%%%%%%%%%%%%%%%%%%%%%%%%%%%%%%%%%%%%%%%%%%%%%%%%%%%%%%%%%%%%
\section*{Glossary}
\noindent
{\em $1d$}: one-dimensional \\
{\em $3d$}: three-dimensional \\
{\em RMS}: root-mean-square\\
{\em CL}: cardiolipin

%%%%%%%%%%%%%%%%%%%%%%%%%%%%%%%%%%%%%%%%%%%%%%%%%%%%%%%%%%%%%%%%%%%%%%%%%%%%%%%%%
\section*{References}

\providecommand{\newblock}{}

\end{document}